\documentclass[10pt]{article}

\usepackage{amsmath}
\usepackage{amssymb}
\usepackage{graphicx}

\usepackage{cite}

\usepackage{color} 

\usepackage{setspace} 
\usepackage[labelfont=bf,labelsep=period,justification=raggedright]{caption}

\makeatletter
\renewcommand{\@biblabel}[1]{\quad#1.}
\makeatother

\date{}

\begin{document}

\begin{flushleft}
{\Large
\textbf{The Effective Fragment Molecular Orbital Method for Fragments Connected by Covalent Bonds}
}
\\
Casper Steinmann$^{1,\ast}$, 
Dmitri G. Fedorov$^{2}$, 
Jan H. Jensen$^{1}$
\\
\bf{1} Department of Chemistry, University of Copenhagen, Universitetsparken 5, DK-2100 Copenhagen, Denmark
\\
\bf{2} NRI, National Institute of Advanced Industrial Science and Technology (AIST), 1-1-1 Umezono, Tsukuba, Ibaraki 305-8568, Japan
\\
$\ast$ corresponding author, E-mail: steinmann@chem.ku.dk
\end{flushleft}

\section*{Abstract}
We extend the effective fragment molecular orbital method (EFMO) into treating fragments connected by covalent bonds. The accuracy of EFMO is compared to FMO and conventional \emph{ab initio} electronic structure methods for polypeptides including proteins. Errors in energy for RHF and MP2 are within 2 kcal/mol for neutral polypeptides and 6 kcal/mol for charged polypeptides similar to FMO but obtained two to five times faster. For proteins, the errors are also within a few kcal/mol of the FMO results. We developed both the RHF and MP2 gradient for EFMO. Compared to \emph{ab initio}, the EFMO optimized structures had an RMSD of 0.40 and 0.44 \AA ~for RHF and MP2, respectively.

\section*{Introduction}
The need to study very large systems in an efficient manner has led to the development of many computational schemes trying to cope with the limitation in computational resources. Linear (or nearly linear) scaling methods have long been of particular interest because they allow, within their respective framework\cite{zhang2003molecular,zhang2003molecular2,jiang2006electrostatic,dahlke2007electrostatically,xie2008variational,xie2007design,soderhjelm2009accurate,soderhjelm2009calculation,beran2009approximating,beran2010spatial,jacobson2011efficient}, large systems to be treated by quantum mechanics. In particular, the use of fragments\cite{Gordon2009review,CRfrag} is very attractive for doing calculations of large systems.

Recently, we developed the effective fragment molecular orbital (EFMO) method\cite{steinmann2010effective}, which builds upon the fragment molecular orbital (FMO) method\cite{kitaura1999fragment,nakano2000fragment,nakano2002fragment,FMOGAMESS,fedorov2007extending,nagata2011mathematical}, and combines it with effective fragment potentials (EFP)\cite{day1996effective,gordon2001effective,ghosh2010noncovalent}. EFMO is different from EFP, FMO and FMO/EFP \cite{nagata2009combined,nagata2011combined} in several ways. For instance, the EFPs are computed on-the-fly from gas phase FMO fragment calculations and used for classical interactions of separated dimers and many-body effects. Extending the earlier work \cite{steinmann2010effective} limited to molecular clusters at the RHF level, we now present the methodology to treat fragments connected by covalent bonds at the MP2 level.

This article is organized as follows. First, we briefly outline the theoretical background of EFMO. We proceed to discuss the change in methodology needed to include fragmentation across covalent bonds in EFMO, including an overview of how fragment bonds are treated. The addition of correlation in EFMO is also presented here. Second, we benchmark the EFMO energy against \emph{ab initio} calculations on three different sets of polypeptides and compare to FMO. We apply our findings to proteins and protein like structures. The quality of the gradient together with timings are also presented here. Water clusters are also briefly revisited. Finally, we summarize our results and discuss future directions.

\section*{Methods}
\subsection*{Theoretical Background}
In FMO, the total two-body (FMO2) non-correlated energy of a system consisting of $N$ fragments (also called monomers) is given as
\begin{equation}
E^\mathrm{FMO2}=\sum_I^N E_I + \sum_{IJ}^N \left(E_{IJ} - E_I  - E_J\right)
\label{eqn:fmo2energy}
\end{equation}
Here $E_I$ ($E_{IJ}$) is the energy of monomer $I$ (dimer $IJ$) in the electrostatic potential (ESP) of the other $N-1$ $(N-2)$ fragments. The momoners converge in the field of ESP, requiring self-consistent charge (SCC) iterations. Dimers converge in the field of ESP of the $N-2$ monomers.

The total non-correlated EFMO energy of a system of $N$ fragments is 
\begin{equation}
E^\mathrm{EFMO} = \sum_I^N E_I^0 + \sum_{IJ}^{R_{I,J}\leq R_\mathrm{resdim}} \left( E_{IJ}^0 - E_I^0 - E_J^0 - E_{IJ}^\mathrm{POL} \right) + \sum_{IJ}^{R_{I,J} > R_\mathrm{resdim}} E_{IJ}^\mathrm{ES} + E_\mathrm{tot}^\mathrm{POL}
\label{eqn:efmoenergy}
\end{equation}
where $E_I^0$ is the gas phase energy of monomer (or fragment) $I$. $E_{IJ}^0$ is the gas phase dimer energy of dimer $IJ$. The second sum in equation~\ref{eqn:efmoenergy} is the pairwise correction to the monomer energy and only applies for dimers separated by a distance less than $R_\mathrm{resdim}$. $E_{IJ}^\mathrm{POL}$ and $E_\mathrm{tot}^\mathrm{POL}$ are the classical pair polarization energy of dimer $IJ$ and the classical total polarization energy, respectively. The final sum over $E_{IJ}^\mathrm{ES}$ is the classical electrostatic interaction energy and applies to dimers separated by a distance greater than $R_\mathrm{resdim}$. The fragment separation distance $R_{I,J}$ was defined previously\cite{steinmann2010effective}. Since EFMO only involves gas phase energy (and gradient) evaluations, only one SCC iteration is required.

In EFMO, the classical terms in the energy expression (equation~\ref{eqn:efmoenergy}) are calculated from expressions in the EFP pertubation expansion of the interaction energy\cite{day1996effective,gordon2001effective}. Based on the converged fragment calculations, EFP parameters are derived on-the-fly completely automatically by computing atom centered monopoles, dipoles, and quadrupoles\cite{stone1981distributed} and dipole polarizability tensors for each electron pair.\cite{minikis2001accurate}

The analytical gradient derived previously\cite{steinmann2010effective} is reformulated for fragments connected by covalent bonds, and also extended to MP2. 

\subsection*{Covalent Bonds}
For fragmentation across covalent bonds, no corrections to the basic equation of EFMO is needed. However, the inclusion of fragmentation across bonds requires a change in the methodology. In this paper, we show how fragmentation is carried out on protein backbones, this methodology is transferable to other systems just as FMO was applied to inorganic systems such as zeolites \cite{fedorov2008covalent} and nanowires \cite{fedorov2009analytic}. 

\begin{figure}[!ht]
\begin{center}
\includegraphics[width=8cm]{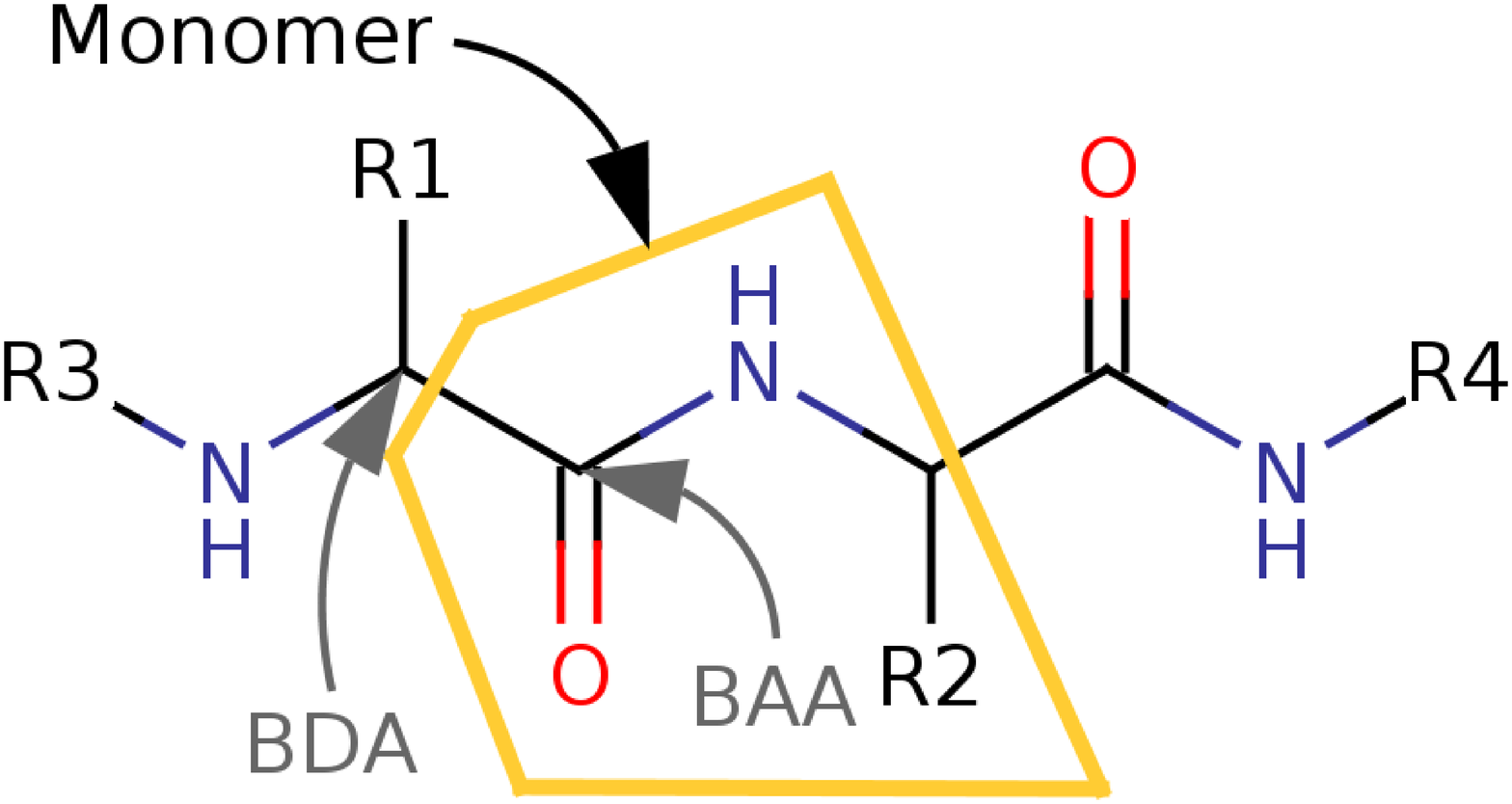}
\end{center}
\caption{
{\bf A model of a backbone in a protein.} The model has side chains (R1 and R2) as well as the continuation of the backbone (R3 and R4). The bond attached atom (BAA) and the bond detached atom (BDA) face each other across the fragmentation point (marked with the yellow line). One fragment is shown within the yellow box.}
\label{fig:modelbackbone}
\end{figure}

In regular FMO, two different schemes of fragmentation is possible. Common to both is that one specifies pairs of atoms which defines fragment boundaries (Figure~\ref{fig:modelbackbone}). Each detached bond is made of a bond attached atom (BAA) and a bond detached atom (BDA). The latter donates an electron to the fragment containing the BAA. One scheme is the hybrid orbital projection (HOP) approach\cite{nakano2000fragment}, which allows full variational treatment of molecular orbitals (MO) across the bond during the fragment SCF. The other is the adapted frozen orbital (AFO) method\cite{fedorov2008covalent,fedorov2009analytic} which freezes the occupied orbital that describes the bond\cite{kairys2000qm}. EFMO uses the latter method, and for completeness we include a discussion of this particular scheme in this work.

\begin{figure}[!ht]
\begin{center}
\includegraphics[width=8cm]{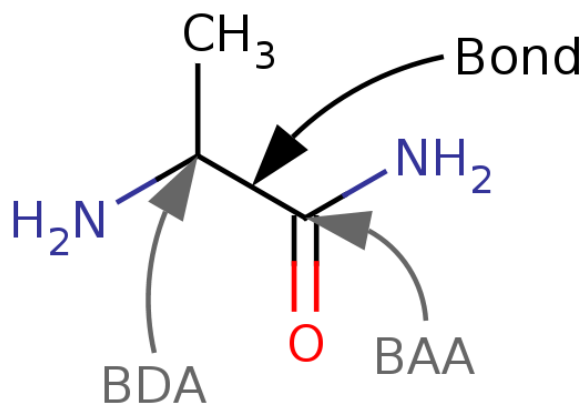}
\end{center}
\caption{
{\bf The current model system used in this study for fragmentation across peptide bonds.} The model is constructed automatically for use with AFO. The central atoms are the bond attached atom (BAA) and the bond detached atom (BDA). The atoms which are connected directly to either the BAA or the BDA are included, capped with hydrogens as necessary.}
\label{fig:modelbond}
\end{figure}

\begin{figure}[!ht]
\begin{center}
\includegraphics[width=8cm]{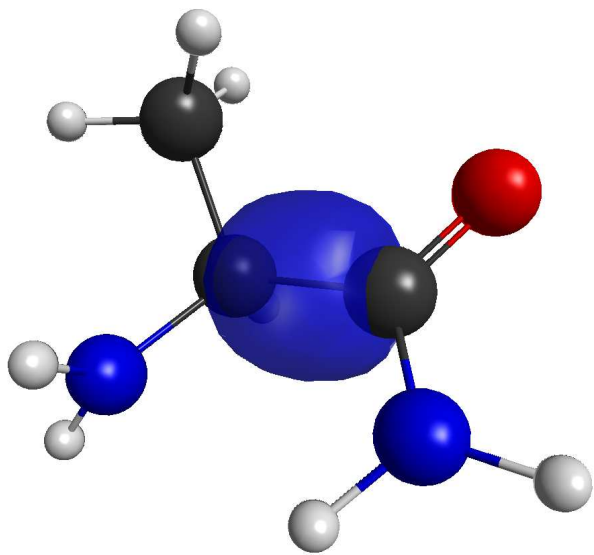}
\end{center}
\caption{
{\bf Special bond orbital for bond 13 in The Trp-cage protein}. The orbital is obtained using RHF/6-31G(d) on a model system (Figure~\ref{fig:modelbond}).}
\label{fig:bondorbital}
\end{figure}

In AFO, a model system around the BAA and BDA is constructed (Figure~\ref{fig:modelbond}). RHF calculations are carried out on this system, followed by an Edminston-Ruedenberg localization\cite{edmiston1963localized}. The occupied orbital which has the largest overlap with the BDA and BAA is identified as the special bond orbital (SBO) shown on Figure~\ref{fig:bondorbital}. This orbital, along with several virtual orbitals on the BDA is stored for later use in monomer and dimer SCF calculations.

For polypeptides, which is the main focus of this study, there is one SBO per pair of BAA and BDA. This SBO is associated with the fragment that contains the BAA. After the computation of all model systems, monomer calculations are done, followed by a Foster-Boys localization, where the SBO is kept frozen, i.e. not allowed to mix with the rest of the orbitals. This leads to a polarizable point in the centroid of the SBO (Figure~\ref{fig:bondorbital}), obtained from the model system across the bond (Figure~\ref{fig:modelbond}). We have thus successfully eliminated the need to manually parametrize the bonds between pairs of fragments.

In the original formulation of EFMO, the electric field arising from a static multipole or induced dipole in fragment $I$ is screened by a Tang-Toennis type expression
\begin{equation}
k(\vec{R},\alpha,\beta) = 1 - \exp \left(-\sqrt{\alpha\beta}|\vec{R}|^2\right)\left(1+\sqrt{\alpha\beta}|\vec{R}|^2\right)
\label{eqn:tangtoennis}
\end{equation}
Here, $\alpha$ and $\beta$ are the screening parameters associated with fragments $I$ and $J$, respectively. The distance parameter $\vec{R}$ is the vector between an induced dipole in fragment $I$ and any of the electric moments in fragment $J$. The above expression is also the default in EFP\cite{day1996effective,gordon2001effective} with the parameters $\alpha=\beta=0.6$. We emphasize that the screening parameters are associated with fragments and not individual polarizable points. 

\subsection*{Correlation}
The introduction of correlation energy in the EFMO method follows previous work in FMO\cite{fedorov2004second,fedorov2005coupled,fedorov2007accuracy}. The total correlated energy of a system of N fragments is given as
\begin{equation}
E = E^\mathrm{EFMO} + E^\mathrm{COR}.
\label{eqn:correlatedenergy}
\end{equation}
Here $E^\mathrm{COR}$ is given as the sum of monomer correlation energies $E_I^\mathrm{COR}$ and pairwise corrections, i.e. 
\begin{equation}
E^\mathrm{COR} = \sum_I^{N} E_I^\mathrm{COR} + \sum_{IJ}^{R_{I,J} < R_\mathrm{corsd}} \left(E_{IJ}^\mathrm{COR} -  E_I^\mathrm{COR} -  E_J^\mathrm{COR}\right),
\end{equation}
where $E_{IJ}^{\mathrm{COR}}$ is the correlation energy of dimer $IJ$. The distance parameter $R_\mathrm{corsd}$ determines whether or not correlation is included for a specific dimer. The value of the parameter is discussed in the computational methodology section below. Note that for the correlation energy any size-extensive post-HF scheme can be used.

\subsection*{Computational Methodology}
All \emph{ab initio} and fragment calculations were carried out in a locally modified version of GAMESS\cite{schmidt1993general}. EFMO was parallelized  with the generalized distributed data interface\cite{fedorov2004new}. In all calculations, the 6-31G(d)\cite{hariharan1973influence,francl1982self,gordon1982self} basis set was employed throughout unless specified otherwise. In all the geometry optimizations, a convergence criterion of $5.0\cdot 10^{-4}$ Hartree / Bohr was used.

The \emph{ab initio} MP2 calculations had their integral accuracy increased to $10^{-12}$ (ICUT=12 in \$CONTRL), SCF convergence criterion was raised from $10^{-5}$ to $10^{-7}$ (CONV=1E-7 in \$SCF) and the MP2 code by K. Ishimura \emph{et. al}\cite{ishimura2006new} with AO integral transformation threshold increased from $10^{-9}$ to $10^{-12}$ (CODE=IMS and CUTOFF=1E-12 in \$MP2) to match what is used in FMO.

For FMO (and EFMO), the AFO scheme was used throughout with the default settings for bond definitions (LOCAL=RUEDNBRG in \$CONTROL and RAFO(1)=1,1,1 in \$FMO). 
The parameters for the electrostatic treatment of dimers $R_\mathrm{resdim}$ and the threshold for the inclusion of correlation effects $R_\mathrm{corsd}$ were both set to 2.0 (RESDIM=2.0 RCORSD=2.0 in \$FMO) unless otherwise specified. The distances are relative to the van-der-Waals radii of atoms (see ref \cite{steinmann2010effective} for details). 
The screening parameter for all fragments are set to 0.1 for fragments with and without the SBO (SCREEN(1)=0.1,0.1 in \$FMO), respectively unless specified otherwise.

The following structures used in this study were taken from previous work by Fedorov et. al.\cite{fedorov2004second,fedorov2007accuracy,fedorov2007fragment} This includes $\alpha$-helices ($\alpha-(\mathrm{ALA})_n$) and $\beta$-sheets ($\beta-(\mathrm{ALA})_n$) of alanine, Chignolin (PDB code: 1UAO) and the Trp-cage (PDB code: 1L2Y). Correlation effects on molecular clusters is carried out by investigating the structures from our previous study\cite{steinmann2010effective}. The crystal structure of the 42 residue protein Crambine (PDB code: 1CRN) is also included and protonated using the PDB2PQR tool\cite{dolinsky2004pdb2pqr,dolinsky2007pdb2pqr}.

The three polypeptides used in this study were constructed by selecting six neutral (at pH = 7) amino acids AIVGLT (P1) and AVSNTL (P2) as well as four neutral and two non-neutral (at pH = 7) residues AVKNTD (P3) and padded with two glycine residues at each end for a total peptide length of 10 residues. The polypeptides were protonated (at pH = 7) using the PDB2PQR tool. P1 had neutral termini (arguments --neutralc --neutraln) while P2 and P3 both had charged termini. For each polypeptide, a conformational search was carried out to locate twenty different structures using the ObConformer tool of the Open Babel package\cite{o2011open,website:openbabel}. They were finally minimized using PM6\cite{stewart2007optimization} in MOPAC\cite{mopac20092008stewart} with a bulk solvent (EPS=80.1).

Only results for two residues per fragment are discussed in detail below, and the results for one residue per fragment are shown in supplimentary materials. We note that because of the large charge transfer in some charged systems the one residue per fragment division leads to very considerable errors. 

When interpreting the accuracy of the results, the following quantities of errors are defined for energies. The error in energy
\begin{equation}
\Delta E^{M,X} = E^M - E^X,
\label{eqn:energydeviation}
\end{equation}
the average deviation of conformers
\begin{equation}
\Delta E^{M,X}_\mathrm{avg} = \frac{1}{N}\sum_I^N \left\{ E_I^M - E_I^X \right\} 
\label{eqn:avgenergydeviation}
\end{equation}
and the mean average deviation (MAD) for conformers
\begin{equation}
\Delta E^{M,X}_\mathrm{MAD} = \frac{1}{N}\sum_I^N \left|E_I^M - E_I^X\right|.
\end{equation}
Here, $M$ is FMO2/HOP, FMO2/AFO or EFMO and $X$ is RHF or MP2. $I$ runs through $N$ conformers of polypeptides. To evaluate the quality of the EFMO gradient, numerical gradients $(\nabla E^\mathrm{num})$ were calculated on $\alpha$-(ALA)$_{10}$ and compared to its analytical counterpart $(\nabla E^\mathrm{ana})$ by the root mean square (rms) deviation of the individual elements
\begin{equation}
\nabla E^{\mathrm{rms}} = \sqrt{\frac{|\nabla E^\mathrm{ana} - \nabla E^\mathrm{num}|}{3 N_{A}}}
\label{eqn:grms}
\end{equation}
and the maximum deviation
\begin{equation}
\nabla E^{\mathrm{max}} = \mathrm{max}\left(\left|\nabla_i E^\mathrm{ana} - \nabla_i E^\mathrm{num} \right|\right).
\label{eqn:gmax}
\end{equation}
$N_{A}$ in equation~\ref{eqn:grms} is the number of atoms in the molecule of interest, $i$ in equation~\ref{eqn:gmax} runs through $3N_{A}$ atomic coordinates.

\section*{Results and Discussion}
\subsection*{Application to Polypeptides}
The performance of EFMO has a critical dependence on the screening parameter (equation~\ref{eqn:tangtoennis}, Figures S1-S3, and Tables S1 and S2) because of the close position of a) induced dipoles located at the centroid of the SBO in one fragment and b) the nearby electrostatic moments and induced dipoles in another (especially, adjacent) fragment. In the following, the screening parameter for all fragments is $\alpha=0.1$ unless otherwise specified.

\begin{figure}[!ht]
\begin{center}
\includegraphics[width=14cm]{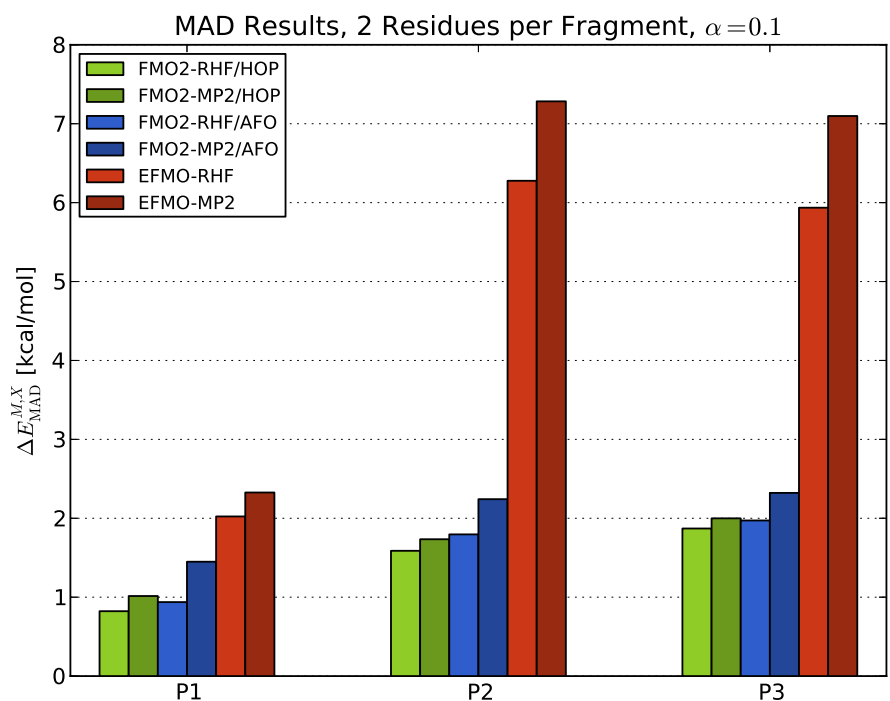}
\end{center}
\caption{
{\bf Mean average deviations of FMO2 and EFMO calculations}. Results are compared to ab initio for conformers of the three polypeptides P1, P2 and P3 using two residues per fragment and the 6-31G(d) basis set. The screening parameter was set to $\alpha=0.1$ for all calculations. Energies in kcal/mol.}
\label{fig:mad2resperfragment}
\end{figure}

Figure~\ref{fig:mad2resperfragment} shows the MAD results obtained for two residues per fragment for all three polypeptides (P1, P2 and P3) using FMO2/HOP, FMO2/AFO and EFMO for both RHF and MP2. For P1, RHF MAD values are 0.82 kcal/mol, 0.94 kcal/mol and 2.02 kcal/mol for FMO2/HOP, FMO2/AFO and EFMO, respectively. The MP2 results yield 1.01 kcal/mol, 1.45 kcal/mol and 2.33 kcal/mol for P1 respectively.

For the charged polypeptide P2, MAD (Figure~\ref{fig:mad2resperfragment}) increases by roughly a factor of two. The factor is about 3 for P3 (from 2.02 kcal/mol to 5.94 kcal/mol for the RHF energy). The inclusion of charged residues results in larger induced dipoles, which has a negative impact on the accuracy of the energy in EFMO. The accuracy of charged systems may be ameliorated by solvent screening.\cite{tomasi2005quantum,molina2003intraprotein,jensen2005prediction}.

\begin{figure}[!ht]
\begin{center}
\includegraphics[width=14cm]{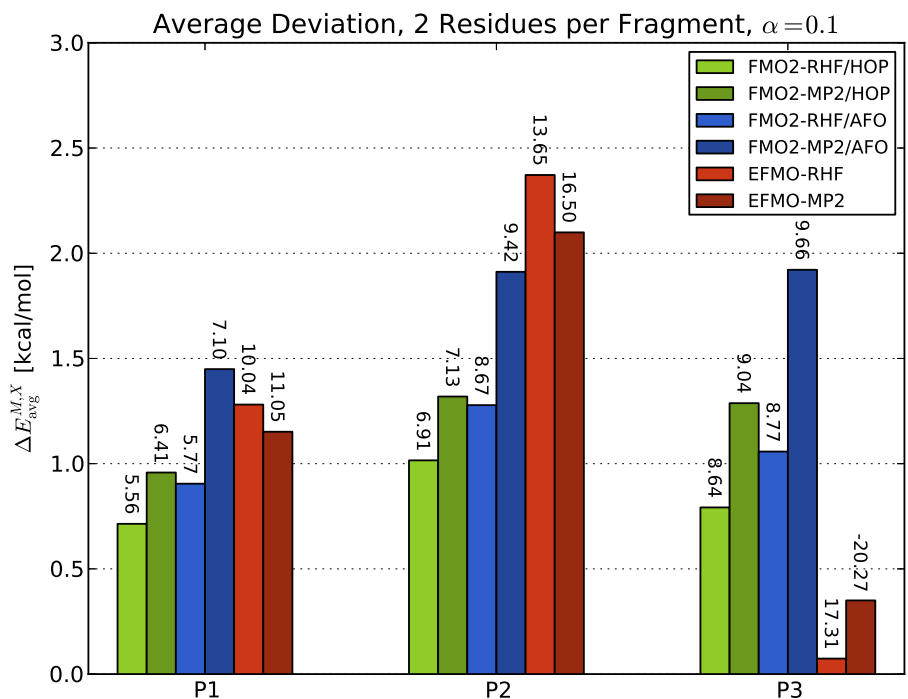}
\end{center}
\caption{
{\bf Average deviations of energy of FMO2 and EFMO calculations compared to RHF and MP2.} All the three polypeptides P1, P2 and P3 using two residues per fragment are shown. Labels on the figure represent the maximum observed deviation. The screening parameter was set to $\alpha=0.1$ for all calculations. Energies are in kcal/mol.}
\label{fig:avg2resperfragment}
\end{figure}

If one considers the average deviation (equation~\ref{eqn:avgenergydeviation} and Figure~\ref{fig:avg2resperfragment}) instead, it is interesting to note that EFMO compares well with FMO2, and the agreement for P3 is perhaps fortuitous (the error is less than 0.5 kcal/mol for EFMO-MP2). The maximum deviations for EFMO, however, are larger in all cases by roughly a factor of two.

For all three peptide ensembles, there is a good correlation between the compactness of the peptide conformation (measured by the radius of gyration) and the error in the energy(see supplimentary information figures S4 to S6). More compact structures place the charged groups closer to the polarizable points at the fragment boundaries resulting in large induced dipoles and errors in the total energy.

\subsection*{Application to Proteins}
The above benchmark of EFMO serves as an initial probe for how the energy behaves for polypeptides as the number of residues per fragment and screening parameters change. Based on those tests, we now apply EFMO to proteins or protein-like structures. The alanine polypeptides are particularly good for studying any systematic error, albeit they are not a representative benchmark for real proteins. 

In Table~\ref{tbl:efmoproteindeviations2res}, deviations in EFMO energy of the various protein structures compared to \emph{ab initio} RHF (MP2) are presented for two residues per fragment with cutoffs $R_\mathrm{resdim}$ and $R_\mathrm{corsd}$ both equal to 2.0. For Chignolin (1UAO), the deviation in energy for EFMO (equation~\ref{eqn:energydeviation}) in RHF (MP2) energy is 1.79 (1.48) kcal/mol, and for FMO2/AFO it is 0.37 (1.38) kcal/mol. For the larger Trp-cage (1L2Y), the EFMO errors are -2.87 (-4.21) kcal/mol and for FMO2/AFO the values are 1.74 (6.35) kcal/mol. The Crambine protein (1CRN) had errors of 15.66 (26.23) kcal/mol for EFMO, which is comparable to the FMO2/AFO results of 3.45 (25.59) kcal/mol. EFMO shows the largest errors of a similar magnitude to FMO2/AFO. Using a 6-31+G(d) basis set on Chignolin, EFMO has the errors of 21.70 (-21.87) kcal/mol. FMO2 did not converge using the default settings.

The results from the $\alpha$-helices and $\beta$-sheets are somewhat more detrimental. With the exception of the RHF EFMO results, the errors are roughly additive for the poly-alanine peptides, so the errors are discussed on a per residue basis. For $\alpha$-helices, the error in energy increase with system size from -2.94 (0.32) kcal/mol for $\alpha-(\mathrm{ALA})_{10}$ to  0.18 (-18.94) kcal/mol for the large $\alpha-(\mathrm{ALA})_{40}$ helix, which corresponds to an average error per residue of 0.29 (0.03) kcal/mol for $\alpha-(\mathrm{ALA})_{10}$ and less than 0.01 (-0.47) kcal/mol for $\alpha-(\mathrm{ALA})_{40}$. The $\alpha$-helices tend to illustrate the case of over-polarization. For $\alpha-(\mathrm{ALA})_{10}$, the total polarization energy is small (-12.89 kcal/mol) but as the system system size increase, so does the total polarization energy (-73.81 kcal/mol) in a non-linear fashion. We note that the MP2 energy for $\alpha-(\mathrm{ALA})_{20}$ and $\alpha-(\mathrm{ALA})_{40}$ increases linearly with system size but the RHF energy does not. The over polarization is also observed for FMO2/AFO, although the MP2 energies are much better (below 2 kcal/mol) which can only be attributed a better wave funtion of the individual fragments and their pairs. The $\beta$-sheets have errors which are lower than in the $\alpha$-alanines the errors are from 0.60 (0.89) kcal/mol to  4.05 (6.46) kcal/mol for $\beta-(\mathrm{ALA})_{10}$ and $\beta-(\mathrm{ALA})_{40}$, respectively. Overall, the average error per residue becomes 0.06 (-0.50) kcal/mol and 0.10 (0.16) kcal/mol for $\beta-(\mathrm{ALA})_{10}$ and $\beta-(\mathrm{ALA})_{40}$, respectively. The $\beta$-sheets are planar and not prone to the same over-polarization (the $\beta-(\mathrm{ALA})_{40}$ has a polarization energy of around 50 kcal/mol).

As noted above, the $\alpha$-helices and $\beta$-sheets illustrate two very different polypeptides. The inaccuracy of EFMO for them is somewhat alleviated by the fact that the errors in energy for Chignolin and the Trp-cage proteins are smaller than the $\alpha$-helices and $\beta$-sheets. The Trp-cage has 20 residues and its error in energy of -2.87 (-4.21) kcal/mol lie around the corresponding $\alpha$-helices and $\beta$-sheets of the same size -2.75 (-9.66) kcal/mol to 1.74 (2.78) kcal/mol, respectively. The same is true for Chignolin.

\subsection*{Gradients and Geometry Optimizations}
A key strength of EFMO over other similar methods\cite{beran2009approximating,beran2010spatial,soderhjelm2009accurate,soderhjelm2009calculation,jacobson2011efficient} is the availability of the gradient. The gradient of FMO2/AFO has been investigated previously for zeolites\cite{fedorov2009analytic} where errors in gradient were found to be $\nabla E^\mathrm{rms}$: $0.2\cdot 10^{-3}$ Hartree/Bohr and $\nabla E^\mathrm{max}$: $1.4\cdot 10^{-3}$ Hartree/Bohr when compared to numerical derivatives (equations \ref{eqn:grms} and \ref{eqn:gmax}) although with a smaller basis set than in this study. It was found, that even though these deviations were present, geometry optimizations did result in satisfactory structures.

In this study, we present an investigation of the EFMO gradient comparing numerical and analytical values for proteins (Table~\ref{tbl:gradientresults}). It has roughly the same accuracy-related issues found for zeolites, specifically around the bond regions where rms and maximum errors for FMO2-RHF/AFO with and without the electrostatic potential is $0.51\cdot 10^{-3}$ Hartree/Bohr, $3.43\cdot 10^{-3}$ Hartree/Bohr and $0.76\cdot 10^{-3}$ Hartree/Bohr and $4.71\cdot 10^{-3}$ Hartree/Bohr, respectively which is on par with what was found for zeolites. The latter result is particularly interesting as it is the FMO2/AFO result on top of which we add the EFP terms to obtain EFMO (equation~\ref{eqn:efmoenergy}).

\begin{figure}[!ht]
\begin{center}
\includegraphics[width=14cm]{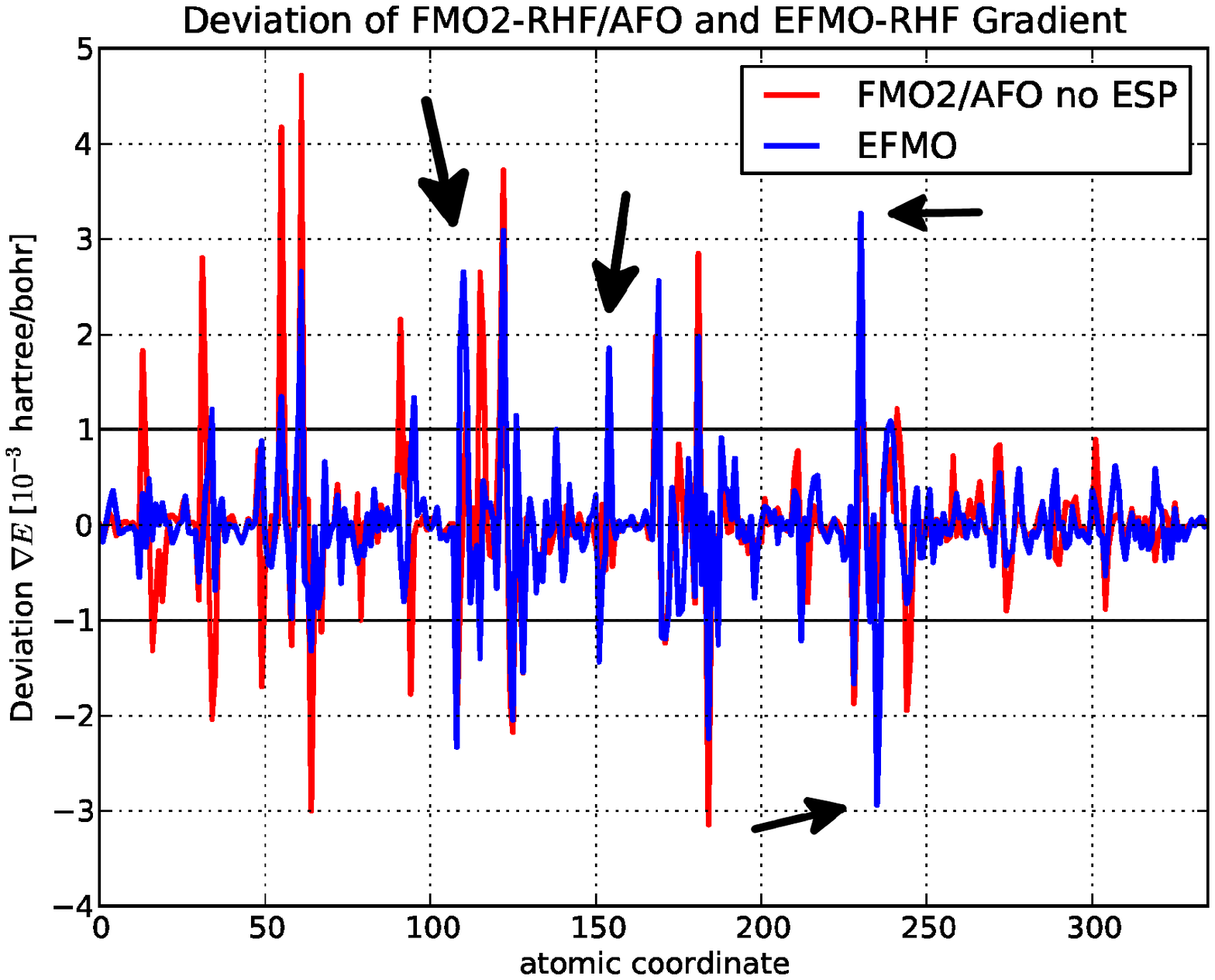}
\end{center}
\caption{
{\bf Deviations of analytic gradient from the numeric gradient for RHF on $\alpha$-(ALA)$_{10}$.} Shown in units of $10^{-3}$ Hartree/Bohr for FMO2-RHF/AFO and EFMO-RHF versus atomic coordinate for the 6-31G(d) basis set.}
\label{fig:efmovsefmonum}
\end{figure}

Several different approaches to tackle the gradient were attempted. The first is the original approach taken for molecular clusters which is to transfer the gradient terms of the induced dipoles $\vec{\mu}^\mathrm{ind}$ to the nearest atom only, in this study named EFMO$_\mathrm{org}$.  This is a clear improvement over the FMO2/AFO (without the ESP) result ($\nabla E^\mathrm{rms}$: $0.73\cdot 10^{-3}$ Hartree/Bohr, $\nabla E^\mathrm{max}$: $3.50\cdot 10^{-3}$ Hartree/Bohr), but some deviations in gradient get worse using EFMO and will be discussed further below. Removing all torque contributions (EFMO$_\mathrm{nt}$) reveals further improvements ($\nabla E^\mathrm{rms}$: $0.68\cdot 10^{-3}$ Hartree/Bohr, $\nabla E^\mathrm{max}$: $3.30\cdot 10^{-3}$ Hartree/Bohr). Another approach, specifically for the induced dipole (EFMO$_\mathrm{nt+pct}$) is to do a percentage based distribution of the induced dipoles based on the distance between two atoms. This only applies if the induced dipole is between two atoms and the gradient is distributed based on a percentage of the entire bond length. This further improves the results, but the improvement ($\nabla E^\mathrm{rms}$: $0.66\cdot 10^{-3}$ Hartree/Bohr, $\nabla E^\mathrm{max}$: $3.27\cdot 10^{-3}$ Hartree/Bohr) reveals that the main source of the error is not due to EFMO (Figure~\ref{fig:efmovsefmonum}), but pertains to approximations in the FMO2/AFO gradient. To make sure that the induced dipoles do not cause major problems, an approach was tried to not evaluate the electric field from the static multipole moments and the induced dipoles, both in the energy and the gradient, of adjacent fragments, that is fragment $I$ covalently bound to fragment $J$ does not induce dipoles in $I$ and vice versa. Results with (EFMO$_\mathrm{nt+pct+adj}$) and without (EFMO$_\mathrm{nt+adj}$) percentage based distribution of induced dipoles are ($\nabla E^\mathrm{rms}$: $0.66\cdot 10^{-3}$ Hartree/Bohr, $\nabla E^\mathrm{max}$: $3.73\cdot 10^{-3}$ Hartree/Bohr) and ($\nabla E^\mathrm{rms}$: $0.66\cdot 10^{-3}$ Hartree/Bohr, $\nabla E^\mathrm{max}$ $3.74\cdot 10^{-3}$ Hartree/Bohr) offer no clear advantage over EFMO$_\mathrm{nt+pct}$ on the RHF level of theory, and consequently MP2 data are not presented.

\begin{figure}[!ht]
\begin{center}
\includegraphics[width=14cm]{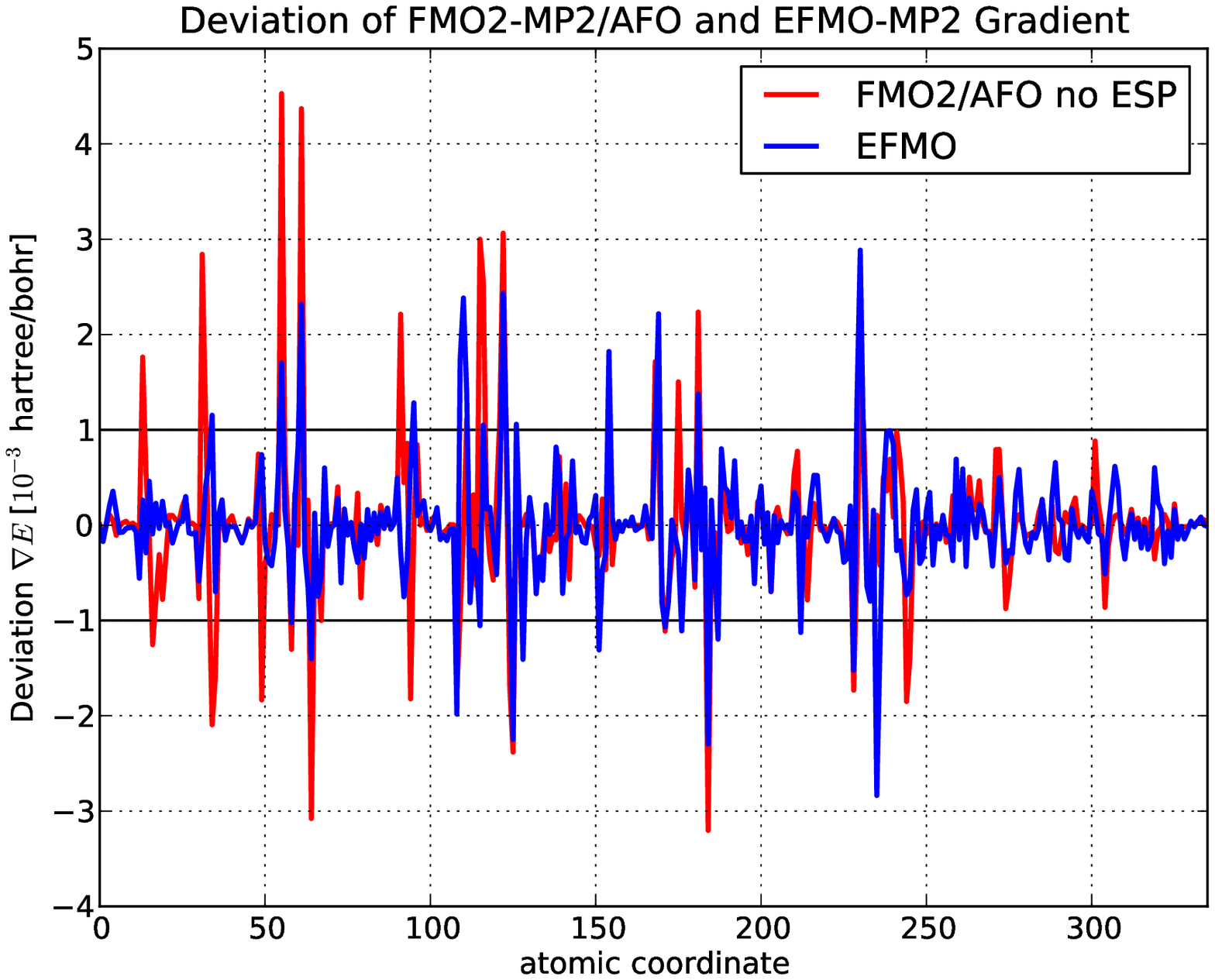}
\end{center}
\caption{
{\bf Deviations of analytic gradient from the numeric gradient for MP2 on $\alpha$-(ALA)$_{10}$.} Shown in units of $10^{-3}$ Hartree/Bohr for FMO2-MP2/AFO and EFMO-MP2 versus atomic coordinate for the 6-31G(d) basis set.}
\label{fig:efmovsefmonummp2}
\end{figure}

From Figure~\ref{fig:efmovsefmonum}, it is clear that EFMO fixes some of the issues that FMO2/AFO has, but evidently creates a few new ones at atom indices 111 (backbone nitrogen), 155 (backbone carbonyl), 231 (backbone nitrogen) and 236 (backbone C$_\alpha$). Common to all is that it is around the bonding region. Evidently, small perturbations in the geometry, specifically around the bonding region, has large implications for the generated EFP parameters. For FMO2-MP2/AFO and EFMO-MP2 (Figure~\ref{fig:efmovsefmonummp2} and Table~\ref{tbl:gradientresults}), the errors in the gradient decrease for the EFMO$_\mathrm{nt+pct}$ methodology ($\nabla E^\mathrm{rms}$: $0.61\cdot 10^{-3}$ Hartree/Bohr, $\nabla E^\mathrm{max}$: $2.89\cdot 10^{-3}$ Hartree/Bohr) while FMO2-MP2/AFO errors are very similar to the corresponding RHF values.

\begin{figure}[!ht]
\begin{center}
\includegraphics[width=8cm]{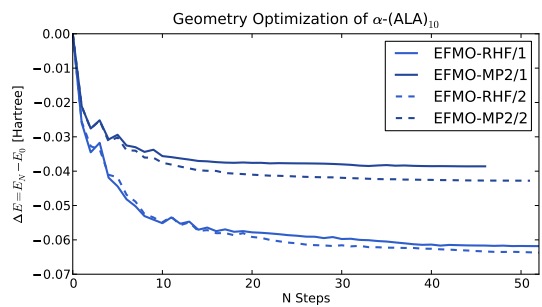}
\end{center}
\caption{
{\bf Convergence of energy as a function of number of geometry steps taken.} Results are from an optimization of $\alpha$-(ALA)$_{10}$ EFMO-RHF and EFMO-MP2 with both one and two residues per fragment calculated using the 6-31G(d) basis set. In all cases, the optimized geometries were optimized to a gradient threshold of $5.0\cdot 10^{-4}$ Hartree/Bohr and all final structures had the lowest energies of all steps taken.}
\label{fig:optimizations}
\end{figure}

Finally, geometry optimizations were carried out for $\alpha$-(ALA)$_{10}$ using the 6-31G(d) basis set and the EFMO$_\mathrm{nt+pct}$ procedure. Figure~\ref{fig:optimizations} shows the improvement in energy as a function of the number of steps taken in a geometry optimization. The obtained optimized structures have the lowest energies when comparing to all the taken steps, even for one residue per fragment. Compared to RHF (MP2) optimized structures, the rms between the optimized structures are 0.40 (0.44) angstrom (EFMO with one residue per fragment did slightly worse). This can be compared to the 0.3 angstrom that was obtained for FMO2-RHF with HOP previously\cite{fedorov2007fragment}.

EFMO offers a gradient whose quality is similar to FMO2/AFO calculations but at a reduced cost. The quality of the FMO2/AFO gradient could be improved if fully analytic derivatives available such as what was done by Nagata \emph{et. al.} for HOP\cite{nagata2010importance,nagata2011fully,nagata2011analytic}. Another improvement can be obtained with an addition of the derivatives of the EFP monopoles (and higher order multipoles) as outlined by Xie \emph{et al}.\cite{xie2008variational} We recommend EFMO$_\mathrm{nt+pct}$ for geometry optimizations of polypeptides.

\subsection*{Molecular Clusters}
Inclusion of correlation in EFMO (equation~\ref{eqn:correlatedenergy}) warrants a new benchmark of the water clusters that was used in the original EFMO paper. In Table~\ref{tbl:waterclusterdeviations}, results for MP2 energies are shown for $R_\mathrm{resdim} = R_\mathrm{corsd} = 2.0 $ for various basis sets. Since there are no covalent bonds, the screening parameter was given its original value of $\alpha=0.6$. In the original EFMO paper, the errors in energy for water clusters were discussed per hydrogen bond (HB) due to EFMO only describing higher order many-body effects for polarization (see ref \cite{steinmann2010effective} for full details), thus, the error is a lack of many-body terms per HB. For EFMO-MP2, only monomer and \emph{ab initio} dimers are considered correlated and the lack of treatment separated dimers gives rise to new errors but we expect these to be small. EFP does include dispersion terms\cite{adamovic2005dynamic}, but these are not included in this work.

The EFMO-MP2/6-31G(d) results deviate by a maximum of 0.78 kcal/mol per HB, which is worse than FMO2-MP2/6-31G(d) which deviates by a maximum of -0.43 kcal/mol per HB. Increasing the basis set shows that the EFMO errors are 0.02 and -0.05 kcal/mol per HB for 6-31+G(d) and 6-31++G(d), respectively. For FMO2, the respective errors are -0.76 and -0.48 kcal/mol. The errors we observe for the larger clusters containing 30, 40 and 50 water molecules are consistent with the smaller 20 water molecule cluster.

\subsection*{Timings}
In our previous study\cite{steinmann2010effective}, EFMO-RHF for molecular clusters were two (five) times faster than the corresponding FMO2 energy (gradient) calculation. In Table~\ref{tbl:walltimemol}, results for Chignolin and the Trp-cage are presented for 5 nodes using 2 cores per node. All timings were carried out on Intel Xeon X5550 CPUs. Here, using EFMO-MP2 instead of EFMO-RHF increases the computation time by roughly a factor of two (from 14.0 minutes to 29.5 minutes for Chignolin using $R_\mathrm{resdim}=2.0$). For FMO2, the same calculation takes 38.5 minutes and 58.6 minutes, respectively. An EFMO-RHF gradient evaluation for Chignolin takes only three minutes longer than the energy, but becomes a five-fold increase when running EFMO-MP2 gradients. The same trends are observed for the Trp-cage. We note a significant speedup when lowering the cutoff distances $R_\mathrm{resdim}$ and $R_\mathrm{corsd}$, especially for the larger Trp-cage. When the cut-off distances go down, the number of \emph{ab initio} dimers decrease. Especially MP2 gradients require much CPU time due to the number of integrals that needs to be transformed\cite{ishimura2006new}.

We note that lowering of the cutoff distances $R_\mathrm{resdim}$ and $R_\mathrm{corsd}$ can have significant impact on the accuracy\cite{FMOGAMESS,fedorov2004second} like we observed for molecular clusters\cite{steinmann2010effective}, however for a modest lowering of the thresholds to $R_\mathrm{resdim}=R_\mathrm{corsd}=1.5$, the energy deviations from \emph{ab initio} are not affected greatly (Table~S3).

\subsection*{Summary}
The effective fragment molecular orbital (EFMO) method is a merger of the effective fragment potential (EFP) method and the fragment molecular orbital (FMO) method and combines the general applicability of the FMO method (for example, to flexible biomolecules) with the speed of the EFP method. In this work, we have introduced new methodology needed to make EFMO work for systems with covalent bonds such as proteins. This, together with the analytical gradient provides an agile tool to treat proteins at a reasonable level of theory. We also showed how to incorporate electron correlation via M\o ller-Plesset perturbation theory.

We made an extensive study on small polypeptides to assess the need for screening when dealing with covalent bonds and found that an additional screening is needed compared to regular EFP. We showed that the deviations in energy on proteins are on par with FMO2 to within a few kcal/mol when using two residues per fragment. For example, Chignolin is reproduced to within 0.1 kcal/mol compared to FMO2. Timings were consistent with our previous work. We obtained two to five times speedup when using EFMO over FMO2 for RHF. The speedup was somewhat lower when employing MP2 gradients, resulting in speedups between 1.6 and 2.3. 

There are many ways in which the EFMO method can be improved and extended, for example, interfacing EFMO with the polarized continuum model (PCM) or the classical dispersion interaction in EFP\cite{adamovic2005dynamic} which would enable us to lower $R_\mathrm{corsd}$ compared to $R_\mathrm{resdim}$, thus speeding up the evaluation of the gradient greatly. Another direction is to follow the multilayer FMO method\cite{fedorov2005multilayer} and the recent frozen domain FMO (FMO/FD) method\cite{fedorov2011geometry}.

FMO has been applied \cite{Sawada2010,CMC2011,Evotec2011} to a number of chemical problems,\cite{Fedorov2012rev} and we expect that EFMO can be a useful method on its own, for example,
in the structure optimization of protein-ligand complexes and other studies related to drug design.
 
\section*{Acknowledgments}
C.S. would like to thank Assistant Professor Hui Li of University of Lincoln-Nebraska and Assistant Professor Lyudmila Slipchenko of University of Purdue for interesting and fruitful discussions. This work was supported by the European Union through the In Silico Rational Engineering of Novel Enzymes consortium and the Danish Research Agency through a Skou Fellowship to J.H.J. Computational resources were provided by a grant from the Danish Center for Scientific Computing (DCSC). D.G.F thanks Prof. Kazuo Kitaura for many fruitful discussions and the Next Generation Super Computing Project Nanoscience Program (MEXT, Japan) and the Computational Materials Science Initiative for financial support. All authors would like to thank Professor Mark S. Gordon for many fruitful discussions, and for suggesting the idea to use
the EFP polarization in the FMO framework.

\bibliography{article}

\section*{Figure Legends}

\newpage

\newpage

\newpage

\newpage

\newpage

\newpage

\newpage

\newpage
\section*{Tables}
\begin{table}[!ht]
\caption{\bf Energy Error compared to \emph{ab initio} calculations on proteins and protein-like structures using two residues per fragment.}
\label{tbl:efmoproteindeviations2res}
\begin{tabular}{lrrrr}
 & \multicolumn{2}{c}{EFMO} & \multicolumn{2}{c}{FMO2/AFO} \\  \hline
 & \multicolumn{2}{c}{$R_\mathrm{resdim} = 2.0 $} & \multicolumn{2}{c}{$R_\mathrm{resdim} = 2.0$} \\
 & RHF & MP2 & RHF & MP2 \\  \hline
$\alpha$-(ALA)$_{10}$ & -2.94 & 0.32 & -0.77 & -0.08 \\
$\beta$-(ALA)$_{10}$  & 0.60 & 0.89 & 0.08 & 0.25 \\
$\alpha$-(ALA)$_{20}$ & -2.75 & -9.66 & -2.30 & -0.53 \\
$\beta$-(ALA)$_{20}$  & 1.74 & 2.78 & 0.22 & 0.71 \\
$\alpha$-(ALA)$_{40}$ & 0.18 & -18.94 & -5.47 & -1.62 \\
$\beta$-(ALA)$_{40}$  & 4.05 & 6.46 & 0.51 & 1.62\\
Chignolin                  & 1.79 & 1.48 & 0.37 & 1.38\\
Trp-cage                  & -2.87 & -4.27 & 1.74 & 6.35 \\
Crambine$^a$              & 15.66 & 26.23 & 3.45 & 25.59 \\  \hline
\multicolumn{5}{l}{$^a$based on an FMO3-MP2/6-31G(d) calculation.}\\
\end{tabular}
\begin{flushleft} We used the 6-31G(d) basis set and $R_\mathrm{resdim} = R_\mathrm{corsd} = 2.0$. In all calculations, the screening parameter $\alpha$ was kept fixed at a value of $\alpha=0.1$. All units in kcal/mol.
\end{flushleft}
\end{table}

\begin{table}[!ht]
\caption{\bf Errors in gradient of EFMO and FMO2/AFO for the $\alpha-$(ALA)$_{10}$ polypeptide using RHF and MP2.}
\label{tbl:gradientresults}
\begin{tabular}{cccccccc}
 & FMO2 & FMO2$^a$ & EFMO$_\mathrm{org}$ & EFMO$_\mathrm{nt}$ & EFMO$_\mathrm{nt+pct}$ & EFMO$_\mathrm{nt+adj}$ & EFMO$_\mathrm{nat+pct+adj}$ \\
\multicolumn{7}{l}{RHF}  \\  \hline
$\nabla E^\mathrm{rms}$ & 0.51 & 0.76 & 0.73 & 0.68 & 0.66 & 0.66 & 0.66 \\
$\nabla E^\mathrm{max}$ & 3.43 & 4.71 & 3.50 & 3.30 & 3.27 & 3.73 & 3.74 \\  \hline
\multicolumn{7}{l}{MP2}  \\
$\nabla E^\mathrm{rms}$ & 0.70 & 0.75 & 0.69 & 1.20 & 0.61 &  & \\
$\nabla E^\mathrm{max}$ & 3.57 & 4.53 & 2.84 & 2.91 & 2.89 &  & \\  \hline
\multicolumn{7}{l}{\footnotesize{ $^a$ No ESP.}}
\end{tabular}
\begin{flushleft}Both RHF/6-31G(d) and MP2/6-31G(d) levels of theory are evaluated. All units in $10^{-3}\;$Hartree/Bohr. The subscripts are: nt for not including torque contributions, pct is a percentage based distribution of the gradient arising from gradient contributins not located on atoms and adj ignores induced dipoles due to neighboring fragments. See text for details.
\end{flushleft}
\end{table}

\begin{table}[!ht]
\caption{\bf Water cluster energy error for EFMO and FMO2 relative to ab initio MP2 (in kcal/mol per hydrogen bond).}
\label{tbl:waterclusterdeviations}
\begin{tabular}{cccc}
 $N_{\ensuremath{\mathrm{H_2O}}}$ & $N_{\ensuremath{\mathrm{HB}}}$ & EFMO & FMO2 \\  \hline
6-31G(d) & & &  \\
  & 31 & 0.63 & -0.43 \\
 20 & 32 & 0.66 & -0.37 \\
  & 29 & 0.78 & -0.38 \\ \hline
6-31+G(d) & & & \\
  & 31 & 0.02 & -0.69 \\
 20 & 32 & 0.01 & -0.67 \\
  & 29 & 0.02 & -0.76 \\ \hline
6-31++G(d) & & & \\
  & 31 & -0.05 & -0.44 \\
 20 & 32 & -0.04 & -0.43 \\
  & 29 & -0.05 & -0.48 \\  \hline
6-31G(d) & & &  \\
30 & 51 & 0.59 & -0.43 \\
40 & 63 & 0.79 & -0.41 \\
50 & 86 & 0.74 & -0.45 \\ \hline
\end{tabular}
\begin{flushleft}Energies are calculated using the 6-31G(d), 6-31+G(d) and 6-31++G(d) basis sets. In all calculations $R_\mathrm{resdim} = R_\mathrm{corsd}= 2.0$ and $\alpha=0.6$
\end{flushleft}
\end{table}

\newpage
\begin{table}[!ht]
\caption{\bf Timings for FMO2 and EFMO energy and gradient calculations on the Trp-cage protein.}
\label{tbl:walltimemol}
\begin{tabular}{cccccc}
  & $R_\mathrm{resdim},R_\mathrm{corsd}$ & $T(\mathrm{RHF})$ & $T(\nabla \mathrm{RHF})$ & $T(\mathrm{MP2})$ & $T(\nabla \mathrm{MP2})$ \\ \hline
\multicolumn{6}{c}{Chignolin} \\
EFMO & 1.0 & 9.6 & 11.1 & 22.8 & 102.8  \\
 & 1.5 & 13.2 & 13.1 & 28.7 & 106.4\\
 & 2.0 & 14.0 & 17.0 & 29.5 & 119.0\\

FMO2 & 2.0 & 38.5 & 59.7 & 58.6 & 149.9$^c$ \\ \hline
\multicolumn{6}{c}{Trp-cage} \\
EFMO & 1.0 & 24.2$^b$ & 23.5 & 42.7 & 161.0  \\
 & 1.5 & 33.7 & 38.3 & 70.7 & 261.6 \\
 & 2.0 & 37.6 & 43.0 & 78.9 & 314.0\\
FMO2 & 2.0 & 100.4 & 187.0 & 142.5 & 408.6$^d$ \\ \hline
\multicolumn{6}{l}{$^a$ tested for both RHF/6-31G(d) and MP2/6-31G(d). All units in minutes.} \\
\multicolumn{6}{l}{CPU utilization was $^b$96\%, $^c$85\% and $^d$91\%. All other were 99\%.}
\end{tabular}
\begin{flushleft}All timings were carried out on 5 nodes containing Intel Xeon X5550 CPUs (10 CPU cores total).
\end{flushleft}
\end{table}


\end{document}